\documentclass[reprint,aps,prl,nofootinbib,superscriptaddress]{revtex4-2}
\usepackage[utf8]{inputenc}
\usepackage{amsmath}
\usepackage{amsfonts}
\usepackage{amsthm}
\usepackage{amssymb}
\usepackage{graphicx}
\usepackage[usenames,dvipsnames]{color}
\usepackage{hyperref}
\usepackage{mathrsfs}
\usepackage{orcidlink}
\usepackage[english]{babel}

\theoremstyle{definition}

\theoremstyle{remark}

\def\actaa{\ref@jnl{Acta Astron.}}      
\def\apj{\textrm{ApJ}}                 
\def\apjl{\textrm{ApJL}}                
\def\aap{\textrm{A\&A}}                

\def\mnras{\textrm{MNRAS}}             
\def\prd{\textrm{Phys.~Rev.~D}}        
\def\prl{\textrm{Phys.~Rev.~Lett.}}    
\usepackage[normalem]{ulem}

\usepackage[usenames]{xcolor}

\begin{document}

\title{Dynamical  pairing in gravitational wave populations}
\author{Fabio Antonini}
\email{antoninif@cardiff.ac.uk}
\affiliation{Gravity Exploration Institute, School of Physics and Astronomy, Cardiff University, 5 The Parade, Cardiff, CF24 3AA, United Kingdom}
\author{Aleksandra Olejak}
\affiliation{Nicolaus Copernicus Astronomical Center, Polish Academy of Sciences, ul. Bartycka 18, 00-716 Warsaw, Poland}
\affiliation{Max Planck Institute for Astrophysics, Karl-Schwarzschild-Str. 1, 85748 Garching, Germany}
\author{Isobel Romero-Shaw}
\affiliation{Gravity Exploration Institute, School of Physics and Astronomy, Cardiff University, 5 The Parade, Cardiff, CF24 3AA, United Kingdom}
\email{Romero-ShawI@cardiff.ac.uk}

\author{{Matthew Mould}}
\affiliation{{Nottingham Centre of Gravity \& School of Mathematical Sciences,\\University of Nottingham, University Park, Nottingham, NG7 2RD, United Kingdom}}

\begin{abstract} 
Binary black holes formed dynamically in  star clusters are often associated with nearly
equal-mass pairing. We show that this expectation is incomplete. The more
general prediction of dynamical assembly is an ordered-pairing relation in which
the two black holes are drawn  independently from a common
one--body mass spectrum and then ordered by mass. 
Once the primary-mass distribution is known, this
relation fixes the conditional companion distribution \(p(m_2\mid m_1)\).
We validate this picture with  cluster simulations. In these models, the ordered-draw relation reproduces the  morphology of the mass-pairing distribution, 
including fixed-\(m_2\) ridges that are inherited from the one--body mass spectrum: features at nearly constant secondary mass that
extend over a  range of primary masses and therefore correspond to
decreasing \(q=m_2/m_1\) as \(m_1\) increases.  Thus, the geometry of the mass
distribution itself becomes a formation-channel diagnostic. Dynamical
pairing maps peaks of the primary-mass spectrum into fixed-\(m_2\)
ridges with predictable amplitude, whereas isolated binary evolution  more naturally 
produces structures at fixed $q$. 
Applying this test to a non-parametric reconstruction of the gravitational-wave
population, we find a tentative enhancement at
\(m_2\simeq30\,M_\odot\) for  \(m_1\gtrsim 40\,M_\odot\) that is broadly compatible with ordered pairing. Finally, we show that dynamical  formation  naturally explains the rapid decline of the
secondary-mass distribution above \(\sim30\,M_\odot\)  
recently reported by
the LIGO--Virgo--KAGRA Collaboration. 
\end{abstract}

\maketitle

\noindent{\emph {Introduction.}} The detections of gravitational waves by the LIGO--Virgo--KAGRA network have begun to reveal the properties of the merging black hole (BH) population \citep{2026arXiv260527223T, theligoscientificcollaboration2026gwtc50methodsidentifyingcharacterizing,2021ApJ...913L...7A,2019ApJ...882L..24A,Abbott:2020gyp,2026arXiv260527226T,2023PhRvX..13a1048A,2025PhRvL.134a1401A,2025arXiv250909123A,2026NatAs..10.1049A,flanagan+2}. A central goal is now to understand the astrophysical origin of these sources, so that the observed masses, spins, and redshifts can be used not only to describe the population, but also to identify and understand the physical environments in which the binaries formed \cite{2023arXiv230401288G, 12025arXiv250602250S, 2022ApJ...941L..39W, 2022ApJ...932L..19B,2020ApJ...891L..27F,2021ApJ...922L...5C,Antoninietal2023, 2024ApJ...962...69F,hussain2024hintsspinmagnitudecorrelationsrapidly, ray2024searchingbinaryblackhole, 2025PhRvD.112b3531A, banagiri2025evidencesubpopulationsmergingbinary, padhyegurjar2026detectionsubpopulationsdelaytimedistribution,2026arXiv260407456G,2026PhRvD.113j3021M}.

One widely discussed formation scenario is dynamical assembly in dense stellar
systems, such as globular clusters, nuclear star clusters, and young massive
clusters \cite{Kulkarni1993,Sigurdsson1993,Banerjee2010,Ziosi2014,Rodriguez2016a,2016ApJ...831..187A,2019MNRAS.487.2947D,2017MNRAS.464L..36A}. In these environments, BHs can pair through few-body
encounters after the formation of the compact-object population. The component
masses are therefore drawn from a reservoir that has already been reshaped by
stellar evolution, retention, mass segregation, and dynamical depletion. A
common expectation is that this process favors nearly equal-mass binaries,
because dynamical interactions preferentially involve the most massive black
holes \cite[e.g.,][]{2016ApJ...824L..12O,2016PhRvD..93h4029R,2017MNRAS.469.4665P,2024ApJ...960...65S}.

Here we show that the preference for massive components in dynamical environments does not imply equal-mass pairing. The companion-mass distribution can instead be governed
by an effective one-body mass spectrum.
  The  consequence is that, once the primary-mass
distribution \(p_1(m_1)\) is known, the model makes a definite prediction for
the conditional companion-mass distribution \(p(m_2\mid m_1)\). The comparison
therefore does not require predicting the full black-hole mass spectrum from
uncertain cluster initial conditions or stellar-evolution physics. Instead, it
provides a null hypothesis to test: does the observed pairing distribution contain
information beyond that already encoded in the effective one-body mass
spectrum?

This question is  related to recent work showing that binary BH mass pairing
contains information beyond the primary-mass spectrum alone
\cite{2020ApJ...891L..27F,2024ApJ...962...69F,2024ApJ...960...65S}.
Ref.~\cite{2020ApJ...891L..27F} introduced random and
mass-ratio-dependent pairing models, and
Ref.~\cite{2024ApJ...962...69F} studied  
whether the data are consistent with primary and secondary masses sharing common features, or with an exchangeable/symmetric description of the two components.
In this work we  ask whether the ordered-draw relation
 arises
as the leading-order outcome of dynamical formation, and test whether a
single-reservoir prediction reproduces the mass-pairing structure of cluster
models and gravitational-wave data.


\bigskip

\noindent {\it Dynamical ordered pairing.}
\label{sec:dynamical_ordered_pairing}
Here we derive and justify the ordered-draw relation between the primary mass distribution and the conditional secondary mass distribution.  The basic idea is that the two
black holes are drawn independently from an effective
 mass distribution incorporating both the initial BH mass function and dynamical effects, and are then ordered so that $m_1\geq m_2$. 

The ordered-draw relation is motivated by the mass dependence of three-body binary formation. Consider an underlying black-hole mass function $n(m)$, defined on $m_{\rm lo}\leq m\leq m_{\rm up}$. For ordered binary components $m_2\leq m_1$, the three-body binary formation rate gives \cite{Antoninietal2023}
\begin{equation}
p_0(m_1,m_2)
\propto
2\int_{m_{\rm lo}}^{m_{\rm up}}
\Gamma_{3b}(m_1,m_2,m_3)\,dm_3 ,
\end{equation}
where the factor of two accounts for ordering. We use the Heggie scaling \cite{1975MNRAS.173..729H}
\begin{equation}
\Gamma_{3b}
\propto
n(m_1)n(m_2)n(m_3)
\frac{(m_1m_2)^4m_3^{5/2}}
{\sqrt{(m_1+m_2)(m_1+m_2+m_3)}} ,
\label{eq:heggie_rate}
\end{equation}
and have assumed equipartition of energies.
The dominant dependence on the two binary components is therefore approximately separable,
$\Gamma_{3b}
\propto
\left[n(m_1)m_1^4\right]
\left[n(m_2)m_2^4\right]
\times
\hbox{weaker mass-sum terms}.
$
Thus, to leading order, three-body formation reweights each object independently according to its own mass. The non-separable dependence on the mass sums produces residual pair correlations \cite[e.g.,][]{2018ApJ...854...41K}, but these are subdominant. We note that this argument is not specific to three-body binary formation and applies to any  dynamical process that does not lead to a strong  correlation between the two component masses.

This motivates the following approximation. We draw two masses independently from an effective BH distribution $p_I(m)$ and then order them:
\begin{equation}
m_1=\max(x,y),
\qquad
m_2=\min(x,y),
\qquad
x,y\sim p_I(m).
\end{equation}
The ordered joint distribution is then
\begin{equation}
p_{\rm od}(m_1,m_2)
=
2p_I(m_1)p_I(m_2),
\qquad
m_{\rm lo}\leq m_2\leq m_1 .
\label{eq:ordered_joint}
\end{equation}
Let
$
F_I(m)=\int_{m_{\rm lo}}^m p_I(m')\,dm'
$
be the cumulative distribution of the effective one--body mass spectrum. The cumulative distribution of the primary mass is
$
G_1(m)=P(m_1\leq m)=F_I^2(m),
$
since the larger of two draws is smaller than $m$ only if both draws are smaller than $m$. Therefore
$
p_1(m)=2F_I(m)p_I(m),
$
and the effective BH distribution is fixed by the primary-mass distribution:
\begin{equation}
F_I(m)=\sqrt{G_1(m)},
\qquad
p_I(m)=\frac{p_1(m)}{2\sqrt{G_1(m)}} .
\label{eq:invert_ordered_draw}
\end{equation}

The conditional companion-mass distribution, $p_{\rm od}(m_2\mid m_1)
=
\frac{p_I(m_2)}{F_I(m_1)} 
$,  becomes
\begin{equation}
p_{\rm od}(m_2\mid m_1)
=
\frac{
p_1(m_2)
}{
2\sqrt{G_1(m_2)}\sqrt{G_1(m_1)}
}.
\label{eq:conditional_m2_ansatz}
\end{equation}
Equivalently, for $q=m_2/m_1\leq 1$,
$
p_{\rm od}(q\mid m_1)
=
{
m_1p_1(qm_1)
}/{
2\sqrt{G_1(qm_1)}\sqrt{G_1(m_1)}
}.
\label{eq:conditional_q_ansatz}
$

Equation~(\ref{eq:conditional_m2_ansatz}) is the central prediction used below. We do not attempt to predict $p_1(m_1)$. Instead, we take the primary-mass distribution of binary BH mergers from either cluster models or GW observations and ask whether the conditional secondary mass distribution is consistent with the ordered-draw prediction implied by $p_1(m_1)$.


We stress that the ordered-draw approximation is a specific expectation for
dynamical pairing, where the component masses are largely inherited from an
effective one-body reservoir. 
For a general ordered joint distribution, $p_1(m_1)$ does not determine $p(m_2\mid m_1)$. The relation above follows only if the two components are approximately independent draws before ordering. This is the behaviour expected for dynamical assembly when the leading interaction rates weight the two objects mainly by their individual masses. It is not expected in general for isolated binary or triple evolution where the two black holes descend from stars born in the same binary and can retain strong correlations imposed by the initial stellar mass ratio, mass transfer, common-envelope evolution, supernova survival, or chemically homogeneous evolution \cite[e.g.,][]{2012ApJ...749...91F,deMink2015,2022arXiv220409061O,2026arXiv260626262C,2022MNRAS.516.1406S}. In the End Matter we show that 
these correlations  preferentially produce structures at approximately fixed \(q\), rather than the fixed-\(m_2\)  expected from  dynamical pairing.

\bigskip
\noindent {\it Comparison to cluster models.} In the Supplementary Material (SM) we show that  the ordered-draw approximation reproduces the results of  analytical calculations where we retain the full non-separable three-body binary formation terms in equation (\ref{eq:heggie_rate}), include binary-single exchanges, and test a simple gravitational-wave merger-time weighting. 

\begin{figure}
\centering
\includegraphics[width=0.47\textwidth]
{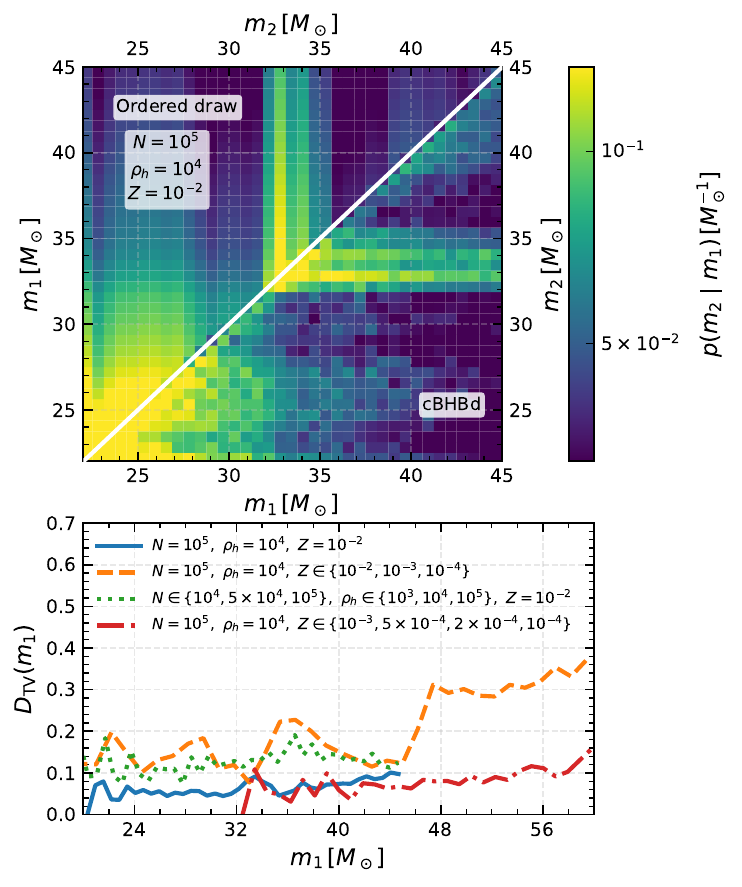}
\caption{Comparison between \texttt{cBHBd} cluster simulations and the ordered-draw reconstruction.
The top two panel show the simulated and reconstructed conditional secondary mass  distributions for $10^4$ realizations of a single
cluster model with \(N=10^5\), \(\rho_{h}=10^4M_\odot~\rm pc^{-3}\), and \(Z=10^{-2}\). 
In the upper left mirrored half the plotted axes are interchanged relative
to the figure coordinates.
The ordered-draw model
captures the main support and secondary-mass structure of the simulated population. The bottom
panel shows \(D_{\rm TV}(m_1)\) for this model and for more heterogeneous populations of clusters with varying cluster properties.
}
\label{fig:1}
\end{figure}

We then carried out a grid of cluster simulations with the fast cluster code \texttt{cBHBd} \cite{AntoniniGieles2020a,2023MNRAS.522..466A,2026arXiv260528088F,2026arXiv260409773C} to test
whether the pairing signatures  persist
in  more detailed dynamical calculations. The models include binary hardening, exchanges,
ejections, binary--binary encounters, long-term cluster evolution, natal kicks, and
realistic initial one--body mass spectra from stellar evolution. We first
consider \(10^4\) random realizations of a single cluster model with initial number of stars
\(N=10^5\), initial half--mass density \(\rho_{h}=10^4M_\odot\rm pc^{-3}\), and metallicity \(Z=0.01\). For this population we
construct the merger distribution in the \((m_1,m_2)\) plane and reconstruct
the ordered-draw prediction from the primary-mass distribution of the same
simulated sample using equation~(\ref{eq:conditional_m2_ansatz}). Figure~\ref{fig:1} shows our main result. The mass ratio distribution at high primary masses is not peaked near one. The ordered-draw model captures
the main support and leading morphology of the simulated distribution,
including a secondary-mass ridge near \(m_2\simeq30\,M_\odot\) extending over
a broad range of primary masses. 
These features are not produced by simple mass ordering, but instead originate from similar structures in the effective reservoir and persist in the conditional distribution because correlations between the component masses are weak.
The agreement is quantified by the conditional
total-variation distance,
\[
D_{\rm TV}
=
\frac{1}{2}
\int
\left|
p_{\rm mod}(m_2\mid m_1)
-
p_{\rm od}(m_2\mid m_1)
\right|dm_2 ,
\]
which is typically \(D_{\rm TV}\lesssim0.1\) for the single-cluster model, indicating that the approximation
 captures most of the conditional probability mass.

The bottom panel of Figure~\ref{fig:1}  shows the comparison for more heterogeneous cluster populations 
(details on  these models are given in the SM).
When models
with different initial cluster masses and densities are combined at fixed
metallicity, the ordered-draw reconstruction remains accurate, with 
\(D_{\rm TV}\lesssim0.1\). 
The agreement degrades when clusters spanning substantially different metallicities, extending up to \(Z=0.01\), are combined, with \(D_{\rm TV}\) reaching \(\sim0.4\) at high primary masses. This occurs because  combining clusters with different black-hole mass ranges and upper-mass cutoffs  introduces sharp boundaries in the joint mass distribution that cannot be fully reconstructed from a single  \(p_1(m_1)\). Nevertheless, as shown in the SM, the ordered-draw approximation still recovers the principal features of \(p(m_2\mid m_1)\), including its ridges and broader structure. Figure~\ref{fig:1} further considers a population spanning a range of  low metallicities, \(Z\le0.001\). Because the corresponding black-hole mass spectra remain broadly similar in this case, the approximation performs substantially better, yielding \(D_{\rm TV}\sim0.1\), comparable to the fixed-metallicity models. In the SM we also consider more accurate cluster Monte Carlo models from Ref.~\cite{2019PhRvD.100d3027R} and demonstrate good agreement with the ordered--draw approximation.
Thus, ordered draws provide a reliable baseline when the constituent clusters have similar black-hole mass spectra, while mixtures of sufficiently different metallicity-dependent populations can produce measurable departures.

Based on these results we conclude that near a peak in the merger rate distribution dynamical formation preferentially leads to $q\sim 1$ as expected. However, at higher masses, there is a preferred
companion-mass scale, so that binaries with increasingly massive primaries are
paired with companions of roughly fixed mass and therefore have decreasing
mass ratios.

\bigskip

\begin{figure}
\centering
\includegraphics[width=0.5
\textwidth]
{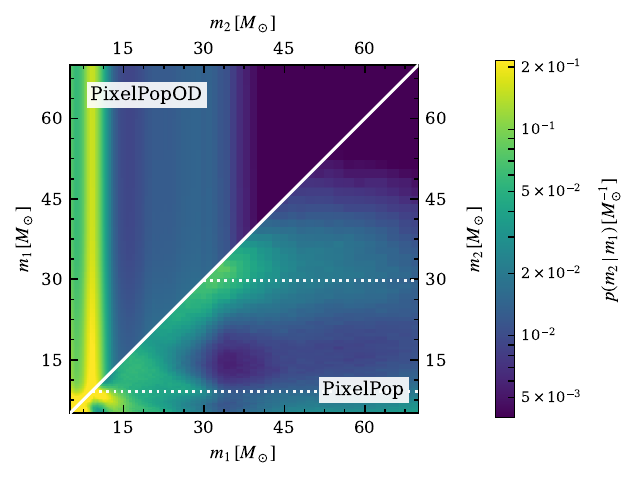}
\caption{
Comparison between the conditional secondary-mass distributions inferred with
\textsc{PixelPop} and \textsc{PixelPopOD}.
The lower-right triangle shows the fully flexible
\textsc{PixelPop} reconstruction.  
The upper-left triangle shows the 
\textsc{PixelPopOD} reconstruction. The dotted horizontal lines indicate the values of 
$
\tilde m_{\rm low}$
and $
\tilde m_{\rm high}$ described in the main text.
}
\label{fig:2}
\end{figure}

\begin{figure}
\centering
\includegraphics[width=0.24
\textwidth]
{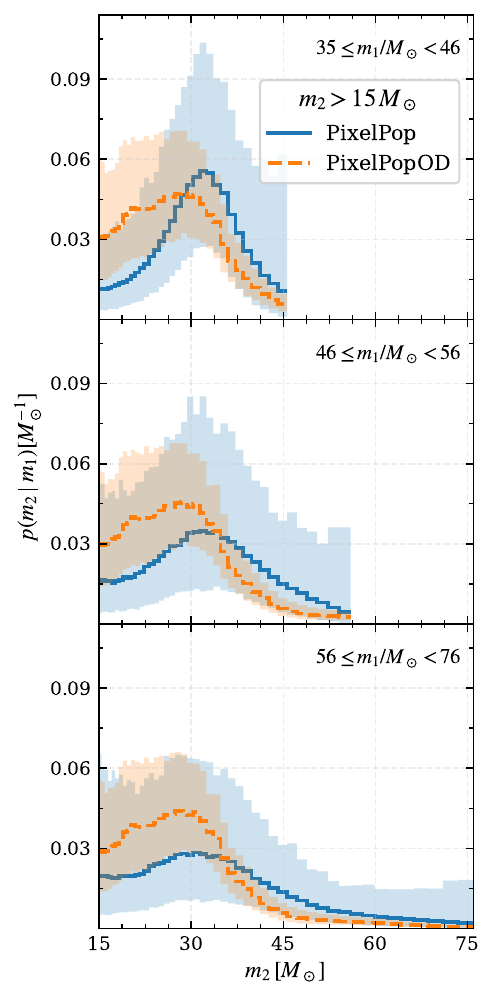}
\includegraphics[width=0.22
\textwidth]
{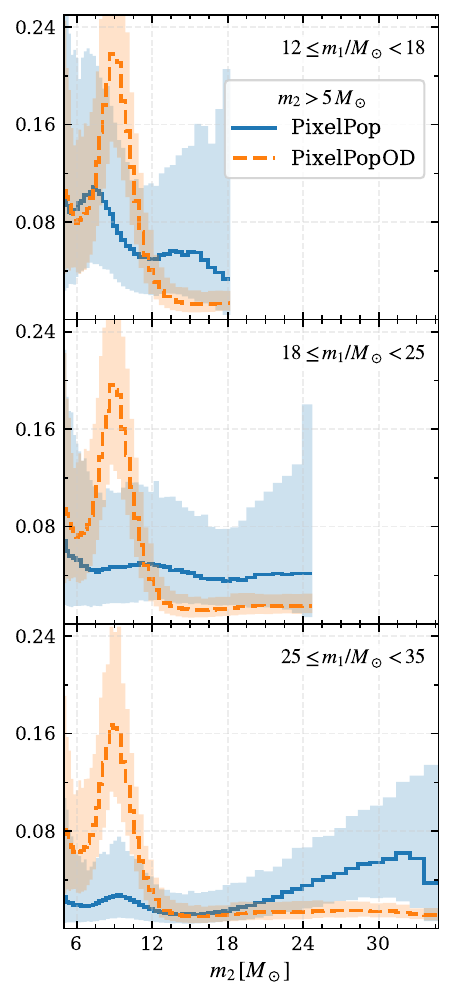}
\caption{
Conditional secondary-mass distributions inferred with \textsc{PixelPop}
 and by \textsc{PixelPopOD},
for the component mass intervals indicated in each panel.
The left and right columns select binaries with both component
masses above $15\,M_\odot$ and $5\,M_\odot$, respectively.
 Lines show posterior medians and shaded regions denote
 $90\%$ credible intervals.
}
\label{fig:3}
\end{figure}

\noindent{\it Application to the gravitational-wave mass distribution.}
\label{sec:data}
We now apply the ordered-draw construction to the nonparametric mass
distribution inferred from the gravitational-wave data.
{We use \textsc{PixelPop} \cite{2025PhRvD.111f3043H,2025PhRvD.111f1305H,2026ApJ..1006L...5A}  to reconstruct the two-dimensional distribution of component masses from the same BH mergers included in \cite{2026arXiv260527226T}. \textsc{PixelPop} is a nonparametric population model that infers the joint merger-rate or probability density over a grid of multidimensional bins. In our analysis, these are \((m_1,m_2)\) pixels, which we then convert into conditional probabilities normalized over each primary-mass column to obtain \(p(m_2\mid m_1)\).}

To test whether the mass pairing inferred from the data is consistent with
ordered drawing we {also} consider a {second} model, \textsc{PixelPopOD}, which
infers a single one-body component-mass spectrum \(p_I(m)\). Binary
components are drawn independently from this spectrum and then ordered, giving
$
p(m_1,m_2)=2p_I(m_1)p_I(m_2)\Theta(m_1-m_2).
$
 We compare this prediction to the fully
flexible \textsc{PixelPop} reconstruction to determine whether the observed secondary
mass structure can be explained by a single ordered-draw reservoir, or whether
additional pairing correlations are required.
{Both models and the population analysis are described in more detailed in the End Matter.}

Figure~\ref{fig:2} shows the posterior-median conditional
distribution from \textsc{PixelPop}. The dotted horizontal lines mark two fixed-secondary-mass
locations predicted using the ordered-draw model.
For each posterior draw of the \textsc{PixelPopOD} mass distribution, we compute the
self-consistent ordered-draw conditional distribution and average it over a
chosen primary-mass interval. We then define the ridge location
\(\tilde m\) as the value of \(m_2\) at which the resulting one-dimensional
profile is maximal. We apply this procedure
to two  selected populations. The first uses the  black-hole
population above \(m_2>5\,M_\odot\), while the second
uses \(m_2>15\,M_\odot\). The latter
selection is therefore not dominated by lower-mass binaries  and isolates the
putative dynamically assembled high-mass population proposed in earlier work
\cite{2023MNRAS.522..466A,banagiri2025evidencesubpopulationsmergingbinary,2026arXiv260614472F}. Repeating the procedure for all posterior draws gives
$
\tilde m_{\rm low}
=
9.1^{+0.9}_{-0.9}\,M_\odot$,
$
\tilde m_{\rm high}
=
29.8^{+4.6}_{-3.3}\,M_\odot,
$
where the quoted values are posterior means with 90\% credible intervals.

Figure~\ref{fig:3} compares the conditional
secondary-mass distributions inferred with \textsc{PixelPop} with that from \textsc{PixelPopOD}  in three primary-mass intervals, separately
for binaries with both component masses above $15\,M_\odot$ and $5\,M_\odot$. 
Since our aim here is to test the
pairing structure of a candidate  population, we condition both
models on the restricted region \((m_1,m_2>m_{\rm cut} )\), and renormalize the conditional distribution over this domain.
The  reason for this is to remove the direct contribution of lower-mass binaries in the \textsc{PixelPopOD} reconstruction and asks whether,
above the chosen threshold, the relative secondary-mass probabilities are
consistent with ordered drawing.


For binaries with both masses above $15\,M_\odot$, ordered drawing
captures the broad concentration of probability near
$m_2\simeq30\,M_\odot$ and its presence across the primary-mass
interval. The predicted distributions are somewhat more
concentrated toward this mass scale than the \textsc{PixelPop} medians,
but they remain  compatible within the large current uncertainties. 
Related features have been noted
previously but with even weaker significance \cite[e.g., Fig.~10 in][]{2024ApJ...960...65S}.

The comparison for binaries with both masses above $5\,M_\odot$
shows larger differences.
Ordered drawing predicts a relatively narrow concentration at
$m_2\lesssim 10\,M_\odot$ across all primary masses. PixelPop, instead, favors
a  broader
conditional distribution across all primary masses. These differences indicate that  in this low mass regime dynamical pairing does not
capture the full  structure  found in the data.

The same ordered-draw geometry also has a direct implication for the
marginal secondary-mass distribution.
In particular, if the high-mass primary tail is approximated by
\(p_1(m)\propto m^\alpha\), with \(\alpha<-1\), then
\(S_I(m)\equiv P_I(>m)\propto m^{\alpha+1}\). In ordered drawing,
\(P(m_1>m)\simeq 2S_I(m)\), so the primary tail has the same
asymptotic slope as \(p_I\). By contrast,
\(P(m_2>m)=S_I^2(m)\propto m^{2(\alpha+1)}\), since both draws must exceed
\(m\), which leads to
\begin{equation}\label{marginal}
p_2(m)\propto m^{2\alpha+1} ~.
\end{equation}  
A
high-mass secondary is  rarer than a high-mass primary because it
requires both objects to be drawn from the high-mass tail of the effective reservoir.

 Equation~\ref{marginal}  provides a simple explanation to the high-mass suppression of
secondaries reported in the GWTC--5 population analysis \cite{2026arXiv260527226T}. The LVK find that above $\sim 30M_\odot$ the
secondary-mass distribution declines more steeply than the primary-mass
distribution, with slopes \(-7.0^{+2.4}_{-2.7}\) and
\(-3.8^{+1.3}_{-1.4}\), respectively.   Taking
the measured primary slope \(\alpha\simeq-3.8\) gives
\(2\alpha+1\simeq-6.6\), consistent with the inferred secondary slope.
Although other explanations have been  proposed \cite[e.g.,][]{2025arXiv250904151T,2026arXiv260527226T,2026PhRvD.113j3021M,OlejakStegmann2026}, our results show that the steeper high-mass decline of \(p_2\) relative to
\(p_1\) can be understood as a generic consequence of dynamical formation and
mass ordering. An immediate implication is that this feature alone is not evidence for
second-generation black holes: while such remnants would preferentially appear
in \(p_1\), a much steeper decline of \(p_2\) relative to \(p_1\) is already expected
from mass ordering (see SM).

\bigskip

\noindent{\it Conclusions.} 
We have introduced the ordered-draw construction as a minimal null hypothesis
for dynamical black-hole pairing. In this picture the two binary components are
drawn independently from an effective one--body mass spectrum, which absorbs the
leading effects of the underlying BH mass function and dynamical
selection, and are then ordered so that \(m_1\ge m_2\).

We have shown that this approximation captures the leading behaviour of
 dynamical cluster calculations. Its accuracy, however, depends on how
well the merging population can be described by a single  one--body
mass reservoir. The ordered-draw prediction works best when the contributing
clusters have similar initial black-hole mass spectra, so that differences
between clusters can be absorbed into one common effective distribution. This
might be a reasonable approximation for cluster populations with a relatively
narrow metallicity range, for example low-metallicity globular clusters, or young massive clusters formed at 
near Solar metallicity.

The ordered draw approximation is expected to degrade for a population of clusters with sufficiently different BH mass distributions. It is also expected to break down
in systems with extended or complex star-formation histories, such as
nuclear star clusters, where multiple metallicity populations that contribute to
the merger sample are formed at different times.  In that case a single ordered-draw reconstruction from the
aggregate primary-mass distribution need not reproduce the full conditional
mass distribution. Nevertheless, even when the  approximation formally breaks
down, fixed-secondary-mass ridges and related regular structures in
\(p(m_2\mid m_1)\) can remain and be used to identify signatures of dynamical formation. Thus, the theoretical picture described in this work can still be used as a baseline for the  interpretation of the observed population.


Our results provide  a simple interpretation of the mass-ratio
distributions found in  \(N\)-body simulations of star clusters \cite[e.g.,][]{Banerjee2017,2019MNRAS.487.2947D,2022A&A...665A..20B,2023MNRAS.526..429A,2025MNRAS.538..639B,2026arXiv260209694B}. These
simulations often show departures from simple equal-mass pairing. In the ordered-draw picture this is not
surprising since dynamical assembly does not require the two components to have
similar masses. Peaks in the one-body reservoir are  mapped into
the two-body distribution as structures at approximately fixed secondary mass.
As a result, massive primaries can be paired with secondaries near a preferred
mass scale, producing mass ratios \(q<1\).

Our results have further implications for how changes in the observed
mass and mass--ratio distributions  are interpreted. Variation in the mass--ratio distribution  with primary mass are often considered as
evidence that different formation channels dominate at different 
masses \cite{ray2026astrophysicaloriginbinaryblack,2025arXiv250915646B,2026arXiv260614472F}. The ordered-draw model shows that this needs not be the case. 
We have shown that the same framework can also explain why the marginal secondary-mass distribution
 declines much more rapidly than the primary distribution at high masses as found in the data \cite{2026arXiv260527226T}.
Moreover, our results have implications for how individual events are interpreted.
The $\sim0.5$ mass--ratio of events such as GW241011 and GW241110 has been argued to be a result of hierarchical mergers in dense environments \cite{2025ApJ...993L..21A}. However, 
we showed that unequal mass binaries away from a peak in the merger rate distribution should be expected to be frequent under standard dynamical pairing of first generation BHs. Thus, caution is needed when interpreting unequal mass ratios as evidence for
a hierarchical-merger origin.

Future
gravitational-wave catalogs will make the  analysis presented here substantially more powerful. Fixed-\(m_2\) ridges corresponding to primary mass features, their contrasts, and their evolution with \(m_1\)  offers a new
direct way to separate pairing signatures inherited from dynamical environments.

\bigskip
\noindent{\it Declaration}. 
After completing this work, we became aware of an independent analysis carried
out concurrently by Ref.~\cite{RoyEtAl_inprep}, who also applied a related
ordered-draw construction to the GWTC--5 data.

\bigskip
\noindent{\it Data and code availability}. The hierarchical
inference code used to produce the results in this paper and the resulting data products from this code and the cluster
simulations are available under
reasonable request.

\bigskip
\noindent{\it Acknowledgment.}
We thank Barry Ginat for providing comments and suggestions on an early version of this manuscript and for serving as LVK internal reviewer. We thank Soumendra Kishore Roy and Thomas Dent for useful comments and suggestions.
FA is supported by the UK’s Science and Technology Facilities Council grants
ST/V005618/1 and UKRI2489.  AO acknowledges support from the Polish National
Agency for Academic Exchange (NAWA) through the Polish Returns Programme,
grant no. BPN/PPO/2025/1/00007/U/00001, and from the National Science Centre,
Poland (NCN), through OPUS 30 grant no. 2025/59/B/ST9/01180.
IMRS acknowledges support from the STFC Ernest Rutherford Fellowship grant number
UKRI2423.
MM is supported by a Research Fellowship from the Royal Commission for the Exhibition of 1851.
This material is based upon work supported by
NSF’s LIGO Laboratory which is a major facility fully funded by
the National Science Foundation, as well as the Science and Technology Facilities Council (STFC) of the United Kingdom, the Max-Planck-Society (MPS), and the State of Niedersachsen/Germany for
support of the construction of Advanced LIGO and construction and
operation of the GEO600 detector. Additional support for Advanced
LIGO was provided by the Australian Research Council. Virgo is
funded, through the European Gravitational Observatory (EGO), by
the French Centre National de Recherche Scientifique (CNRS), the
Italian Istituto Nazionale di Fisica Nucleare (INFN) and the Dutch
Nikhef, with contributions by institutions from Belgium, Germany,
Greece, Hungary, Ireland, Japan, Monaco, Poland, Portugal, Spain.
KAGRA is supported by Ministry of Education, Culture, Sports, Science and Technology (MEXT), Japan Society for the Promotion of
Science (JSPS) in Japan; National Research Foundation (NRF) and
Ministry of Science and ICT (MSIT) in Korea; Academia Sinica
(AS) and National Science and Technology Council (NSTC) in Taiwan. This research has made use of data or software obtained from the Gravitational Wave Open Science Center (gwosc.org), a service of the LIGO Scientific Collaboration, the Virgo Collaboration, and KAGRA. The authors are grateful for computational resources provided
by Cardiff University and supported by STFC grant
ST/V005618/1.

\bigskip

\appendix*
\onecolumngrid
\begin{center}
\onecolumngrid
\section{\large ---End Matter---}
\end{center}
\twocolumngrid

\subsection{Isolated binary evolution}

Mass-transfer interactions in the isolated binary evolution channel may produce merging binary BHs spanning a broad range of mass ratios, \(q\sim0.4-1.0\), but with characteristic features and preferences arising from the orbital evolution during mass-transfer phases \citep{2021AA...651A.100O, vanSon2022, MacLean2026, Gallegos-Garcia2026}. In particular, two broad characteristic peaks have been identified: a population of nearly equal-mass binary BH mergers and a second population at more unequal mass ratios, typically \(q\sim0.5\text{--}0.7\), associated with different mass-transfer subchannels \citep{Olejak2024, Gallegos-Garcia2026}. Nearly equal-mass  binary BHs are predicted to arise from systems undergoing early mass transfer (Case A) \citep{Briel2026}, as well as from systems experiencing common-envelope evolution. In contrast, the unequal-mass peak at \(q\sim0.5\text{--}0.7\) can arise when the first mass-transfer episode leads to mass-ratio reversal and significant orbital widening, followed by a later Case B mass-transfer episode onto the BH \citep{2021AA...651A.100O}. The exact location of this peak may vary, shifting toward more equal mass ratios if BH natal kicks are stronger or if the material lost from the binary carries away more specific angular momentum than assumed under isotropic re-emission \citep{Olejak2024}. The relative prominence and detailed properties of these populations also depend on the stability and degree of conservativeness of mass transfer \citep{vanSon2022, Olejak2024, MacLean2026, Gallegos-Garcia2026}.

\begin{figure}
\centering
\includegraphics[width=0.23
\textwidth]
{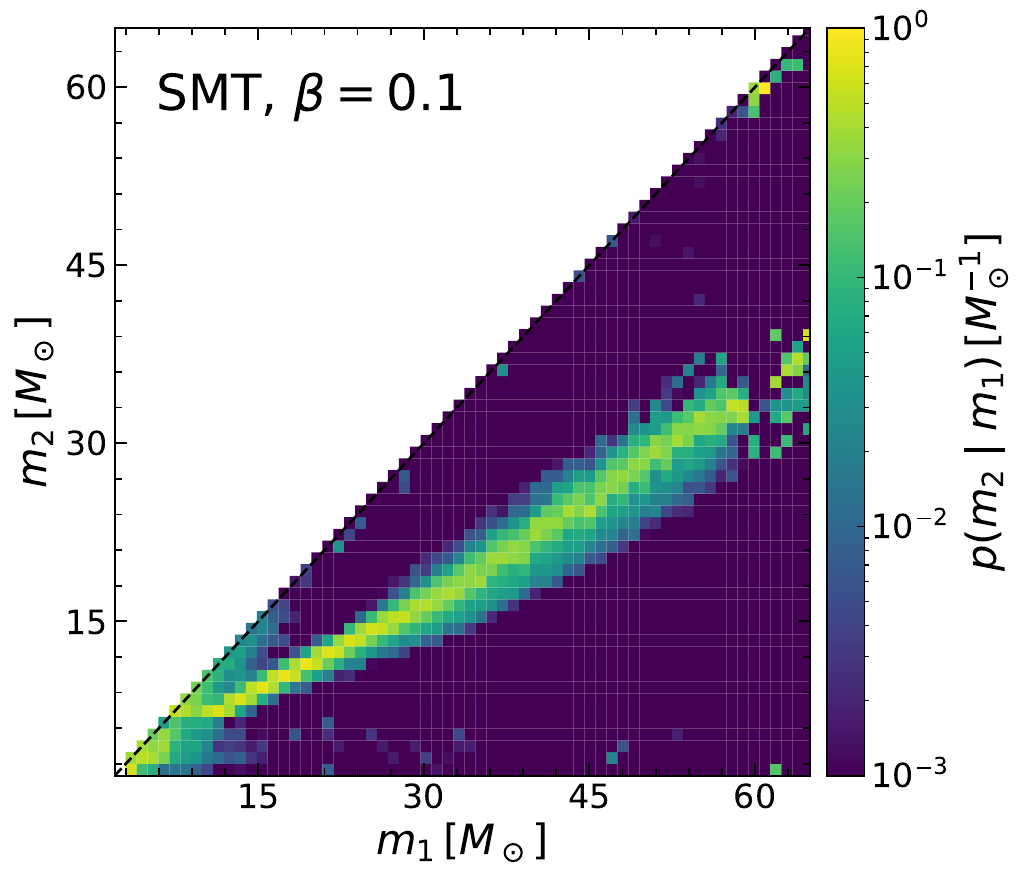}
\includegraphics[width=0.23
\textwidth]
{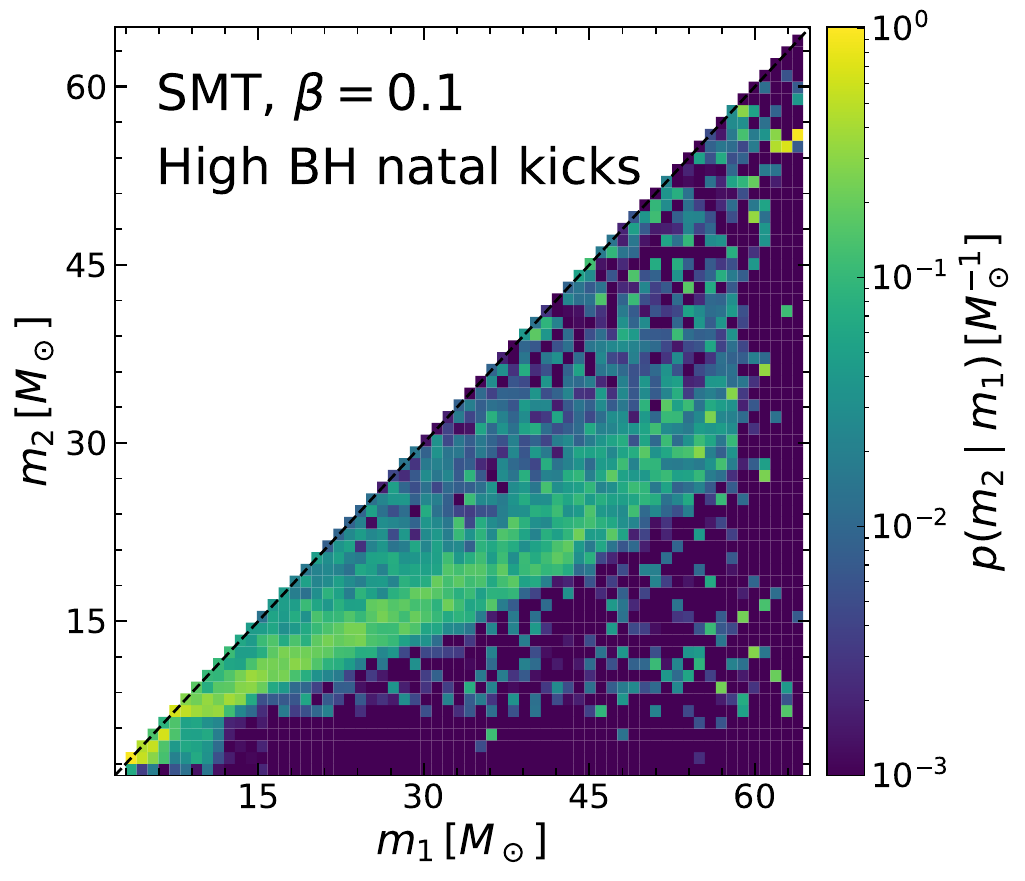}
\includegraphics[width=0.23
\textwidth]
{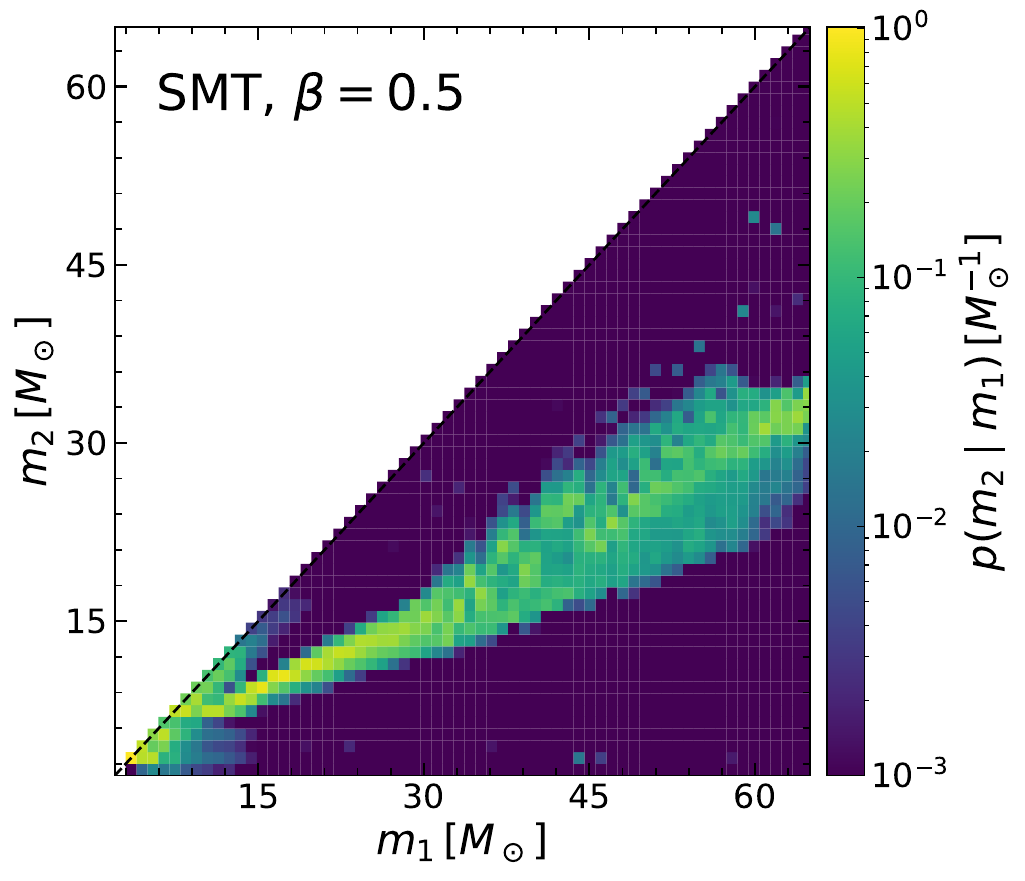}
\includegraphics[width=0.23
\textwidth]
{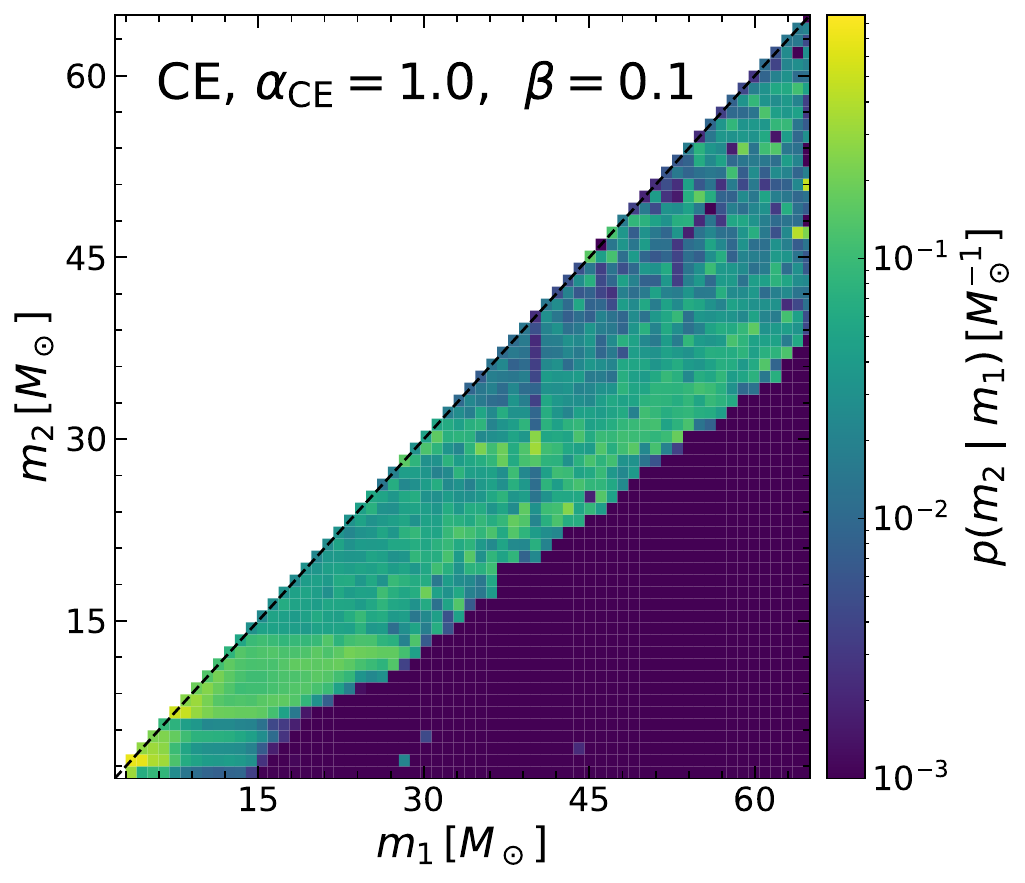}
\caption{
Binary BH populations formed through isolated binary evolution and merging at redshifts $z<0.3$. We show the standard model, dominated by the stable mass transfer (SMT) channel with a non-conservative first mass-transfer phase $\beta = 0.1$ (upper left); a variation with high BH natal kicks (upper right); a variation with a more conservative ($\beta = 0.5$) first mass-transfer phase (lower left); and a model adopting different mass-transfer stability criteria, dominated by common-envelope (CE) evolution during the second mass-transfer phase (lower right). Each primary-mass column is normalized independently.}
\label{fig:iso}
\end{figure}

In Figure~\ref{fig:iso}, we present results for four models of binary BH populations formed through isolated binary evolution and merging at redshifts $z<0.3$. The populations were generated with the {\tt StarTrack} population-synthesis code \citep{Belczynski2008, Belczynski2020}. Our standard model (upper left), presented and described in \citet{OlejakStegmann2026}, favors formation through stable mass transfer (SMT) and assumes highly non-conservative mass transfer with a mass-accretion efficiency during the first mass transfer of $\beta=0.1$ (i.e., 90\% of the transferred mass is lost from the system). The high-kick variant (upper right) adopts the same binary-evolution assumptions but allows for substantially larger BH natal kicks by removing the fallback-dependent reduction in kick magnitude. We also show a model with a more conservative ($\beta=0.5$) first mass-transfer phase (lower left), and a model adopting different mass-transfer stability criteria, in which binary BH formation is dominated by systems undergoing common-envelope (CE) evolution during the second mass-transfer phase (lower right). All four models adopt a high pair-instability cutoff, allowing the formation of first-generation BHs up to $90\,M_{\odot}$\citep{Belczynski2020b}.

As shown in Figure \ref{fig:iso}, mass-transfer evolution produces distinct structures in the binary BH mass distribution associated with characteristic mass ratios. In particular, the SMT models exhibit a transition in the preferred mass ratios between lower- and higher-mass systems. At lower primary masses, $m_1<20\,M_{\odot}$, the models predict a broad mass-ratio distribution, with a preference for nearly equal-mass binaries. At higher primary masses, $m_1>20\,M_{\odot}$, a prominent unequal-mass component emerges at $q\sim0.5$--$0.7$. Larger BH natal kicks enhance the contribution of nearly equal-mass mergers at high masses while also broadening the range of mass ratios represented in the population. Nevertheless, the characteristic mass-dependent trends produced by the underlying binary evolution remain visible. The CE-dominated model presented here preferentially produces nearly equal-mass binary BH mergers across the entire primary-mass range. However, the CE mass-ratio distribution is more sensitive than that of the SMT subchannel to, e.g., the assumed core-collapse remnant-mass prescription \citep{2022MNRAS.516.2252O}, and can shift toward more unequal masses under different assumptions \citep{Gallegos-Garcia2026}. Overall, the structures generated by isolated binary evolution approximately follow lines of constant \(q\), rather than the fixed-\(m_2\) ridges expected from  dynamical formation.

\subsection{Gravitational-wave population analysis details}

We follow the analyses of Ref. \cite{2026arXiv260527226T}, including the same 259 BH mergers with false-alarm rates $<1$\,yr$^{-1}$. We use the \textsc{NRSur7dq4} \cite{2019PhRvR...1c3015V} parameter estimates where available, otherwise \textsc{SEOBNRv5PHM} \cite{2023PhRvD.108l4037R} for events in the fourth observing run and \textsc{IMRPhenomXPHM} \cite{2021PhRvD.103j4056P} for remaining events (due to issues with other analyses \cite{2026arXiv260915827M}). To prevent large Monte Carlo variance $V$ due to finite sampling effects \cite{2023MNRAS.526.3495T,2025arXiv250907221H}, we smoothly penalize the population likelihood by a factor $e^{-100(V-1)^2}$ when $V>1$.

We use \textsc{PixelPop} \cite{2025PhRvD.111f3043H,2025PhRvD.111f1305H} to model the BH merger population, specifically the variant using an intrinsic conditional autoregressive (ICAR) prior \cite{2026ApJ..1006L...5A}. Over BH masses, we place 100 log-uniform bins between 5$M_\odot$ and 150$M_\odot$. For the standard analysis, this results in 5050 bins the in the $(m_1,m_2)$ plane subject to $m_2 \leq m_1$, with bins along the diagonal having half the area. We infer the probability density in each of these bins. The \textsc{PixelPopOD} analysis instead infers the probability densities in each of the 100 bins for the one-body mass distribution $p_I(m)$ directly, then converts to $p_\mathrm{od}(m_1,m_2)$ following Eq.~\eqref{eq:ordered_joint}. In both cases, alongside the BH mass distribution we also use one-dimensional \textsc{PixelPop} models to infer the probability density for the effective spin $\chi_\mathrm{eff}$ \cite{2026arXiv260527223T} in 100 uniform bins over $-1<\chi_\mathrm{eff}<1$ and the merger rate per unit comoving volume and source-frame time over 100 redshift bins $0<z<1.5$. All \textsc{PixelPop} posteriors are marginalized in closed form over the ICAR precision parameter using an improper log-uniform prior.

\clearpage
\newpage

\onecolumngrid
\section{\label{sec:SupMat} Supplementary Material}
\setcounter{equation}{0}
\setcounter{figure}{0}
\renewcommand{\theequation}{S\arabic{equation}}
\renewcommand{\thefigure}{S\arabic{figure}}
\renewcommand{\thetable}{S\arabic{table}}

\subsection{Validation of the ordered-draw approximation with simple analytical models}
\label{sec:supp_ordered_draw}

In the main text we use the ordered-draw relation as a baseline prediction for dynamical pairing. Here we test how accurately this approximation reproduces a more explicit dynamical calculation that includes the effect of exchanges in binary-single interactions, and gravitational wave radiation. Then we consider a full dynamical model of star clusters that takes into account all relevant dynamics. The logic is to compare the conditional secondary mass distribution produced by the dynamical calculation with the ordered-draw prediction reconstructed from the \emph{same} primary-mass distribution. In this way, changes in the effective one--body mass spectrum are absorbed into the ordered-draw prediction, and the comparison isolates residual correlations between the two binary components.

We assume an underlying black-hole mass function
\begin{equation}
n(m)\propto m^\alpha ,
\end{equation}
defined on the interval $m_{\rm lo}\leq m\leq m_{\rm up}$, with $m_{\rm lo}=5M_\odot$ and $m_{\rm up}=50M_\odot$. For a binary with ordered component masses $m_2\leq m_1$, the  binary formation distribution is obtained by integrating the three-body binary formation rate over the mass of the third object:
\begin{equation}
\tilde p_0(m_1,m_2)
=
2
\int_{m_{\rm lo}}^{m_{\rm up}}
\Gamma_{3b}(m_1,m_2,m_3)\,dm_3 ,
\qquad m_2\leq m_1 .
\end{equation}
The factor of two accounts for the ordering of the two binary components. We use the Heggie three-body scaling \cite{1975MNRAS.173..729H}
\begin{equation}
\Gamma_{3b}(m_1,m_2,m_3)
\propto
n(m_1)n(m_2)n(m_3)
\frac{(m_1m_2)^4m_3^{5/2}}
{\sqrt{(m_1+m_2)(m_1+m_2+m_3)}} .
\end{equation}
After normalization,
\begin{equation}
p_0(m_1,m_2)
=
\frac{\tilde p_0(m_1,m_2)}
{
\int_{m_{\rm lo}}^{m_{\rm up}}dm_1
\int_{m_{\rm lo}}^{m_1}dm_2\,
\tilde p_0(m_1,m_2)
},
\end{equation}
this defines the initial no-exchange joint distribution of dynamically formed binaries.

We then model subsequent binary-single encounters as an exchange kernel acting on this distribution. Consider a binary with component masses $a\geq b$ interacting with an interloper of mass $c$. For a fixed binary $(a,b)$, the interloper mass is drawn from the encounter-weighted distribution \cite{2023MNRAS.522..466A}
\begin{equation}
g(c\mid a,b)
=
\frac{
n(c)c^{1/2}\sqrt{a+b+c}
}{
\int_{m_{\rm lo}}^{m_{\rm up}}
n(c')c'^{1/2}\sqrt{a+b+c'}\,dc'
}.
\end{equation}
This expression is the conditional interloper distribution for a binary that undergoes an encounter. The full encounter rate contains an additional factor proportional to
$
\sqrt{(a+b)ab},
$
but this factor is independent of $c$ for fixed $(a,b)$ and therefore cancels in the normalized distribution $g(c\mid a,b)$. In the present validation calculation we fix the number of encounters, $N_{\rm enc}$, rather than assigning a mass-dependent encounter number to each binary.

For each binary-single encounter we use the prescription from Ref.~\cite{2021PhRvX..11c1020G} to assign the probability that each pair remains bound after the resonant interaction. We define
$
W_{ab}
=
\frac{a^4b^4}{(a+b)^{5/2}},
$
$
W_{ac}
=
\frac{a^4c^4}{(a+c)^{5/2}},
$ and
$
W_{bc}
=
\frac{b^4c^4}{(b+c)^{5/2}} .
$
The probabilities that the final binary is $(a,b)$, $(a,c)$, or $(b,c)$ are
\begin{equation}
P_{ab}
=
\frac{W_{ab}}{W_{ab}+W_{ac}+W_{bc}},
\qquad
P_{ac}
=
\frac{W_{ac}}{W_{ab}+W_{ac}+W_{bc}},
\qquad
P_{bc}
=
\frac{W_{bc}}{W_{ab}+W_{ac}+W_{bc}} .
\end{equation}
The term $P_{ab}$ corresponds to no exchange, while $P_{ac}$ and $P_{bc}$ correspond to exchange outcomes.

We  write the exchange step as an integral operator. Let
\begin{equation}
\delta_{\rm od}(m_1,m_2;x,y)
=
\delta\!\left[m_1-\max(x,y)\right]
\delta\!\left[m_2-\min(x,y)\right]
\end{equation}
denote the ordered two-body delta function. The outcome kernel for an encounter between $(a,b)$ and $c$ is
\begin{align}
K(m_1,m_2\mid a,b,c)
=
P_{ab}\,\delta_{\rm od}(m_1,m_2;a,b)
+
P_{ac}\,\delta_{\rm od}(m_1,m_2;a,c)
+
P_{bc}\,\delta_{\rm od}(m_1,m_2;b,c) .
\end{align}
The binary distribution after one encounter is 
\begin{align}
p_{k+1}(m_1,m_2)
=
\int_{m_{\rm lo}}^{m_{\rm up}}da
\int_{m_{\rm lo}}^{a}db\,
p_k(a,b)
\int_{m_{\rm lo}}^{m_{\rm up}}dc\,
g(c\mid a,b)
K(m_1,m_2\mid a,b,c) ,
\end{align}
with $p_k(m_1,m_2)$ the distribution after $k$ encounters. Starting from $p_0(m_1,m_2)$ and iterating gives the equation:
\begin{equation}
p_{\rm ex}(m_1,m_2;N_{\rm enc})
\equiv
p_{N_{\rm enc}}(m_1,m_2).
\end{equation}
The calculation evolves the probability distribution itself. Thus each binary  contributes probability weight to all three possible outcomes, with weights $P_{ab}$, $P_{ac}$, and $P_{bc}$.

From the final distribution we compute the primary-mass distribution
\begin{equation}
p_{1,\rm ex}(m_1)
=
\int_{m_{\rm lo}}^{m_1}
p_{\rm ex}(m_1,m_2;N_{\rm enc})\,dm_2 ,
\end{equation}
and the corresponding conditional companion mass distribution
\begin{equation}
p_{\rm ex}(m_2\mid m_1)
=
\frac{
p_{\rm ex}(m_1,m_2;N_{\rm enc})
}{
p_{1,\rm ex}(m_1)
}.
\end{equation}
This explicit conditional distribution is then compared to the ordered-draw prediction reconstructed from the same primary-mass distribution, $p_{1,\rm ex}(m_1)$.

The exchange calculation above describes the mass distribution of dynamically assembled binaries after a fixed number of encounters. 
The  case $N_{\mathrm{enc}}=0$, which we also consider,  corresponds to the population produced directly by three-body binary formation, before any subsequent binary-single processing. 
Thus,  this model tests whether the mass correlations generated by three-body binary formation can be reproduced by ordered draws from an effective one-body mass distribution. 

To test whether gravitational-wave inspiral selection can introduce significant correlations between the binary components, we also consider a simple merger-time weighting.
For fixed semi-major axis and eccentricity, the Peters merger time scales as \cite{Peters1963}
\begin{equation}\label{gwtime}
t_{\rm GW}
\propto
\frac{a^4}{m_1m_2(m_1+m_2)}
f(e).
\end{equation}
For highly eccentric binaries, writing $\ell^2=1-e^2$, one has $f(e)\propto \ell^7$. If the eccentricity distribution is approximately thermal, then $\ell^2$ is uniformly distributed. The probability that a binary merges within a fixed time therefore scales approximately as
\cite{AntoniniGieles2020a}
\begin{equation}
P
\propto
\left[m_1m_2(m_1+m_2)\right]^{2/7},
\end{equation}
at fixed semi-major axis.

We implement this effect by reweighting the explicit dynamical distribution according to
\begin{equation}
p_{\rm GW}(m_1,m_2)
\propto
p_{\rm ex}(m_1,m_2)
\left[m_1m_2(m_1+m_2)\right]^\eta ,
\end{equation}
 The value $\eta=2/7$ corresponds to the thermal-eccentricity merger-probability scaling above. We set  $\eta=1$ as a deliberately stronger, $1/t_{\rm GW}$-like weighting. 

After applying this weight, we renormalize the distribution and compute its primary-mass distribution,
$
p_{1,\rm GW}(m_1)$.
We then compare
$p_{\rm GW}(m_2\mid m_1)$
to $p_{\rm od}(m_2\mid m_1;p_{1,\rm GW}) $.

\begin{figure*}
\centering
\includegraphics[width=0.95\textwidth]
{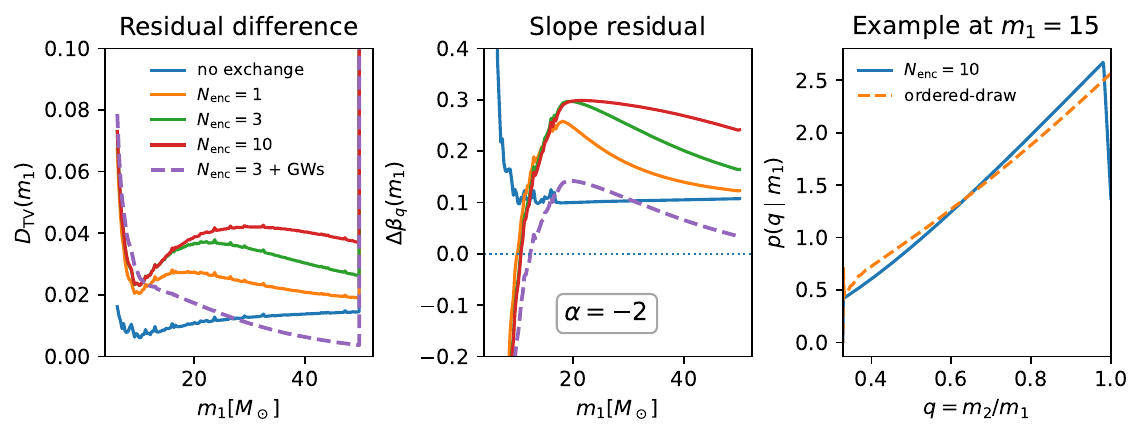}

\includegraphics[width=0.95\textwidth]
{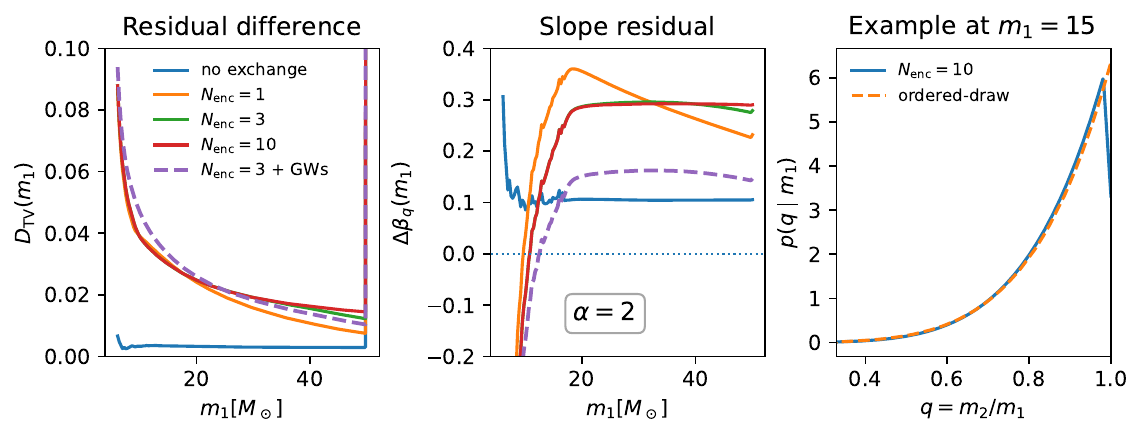}
\caption{
Validation of the ordered-draw approximation for two underlying black-hole mass functions, $n(m)\propto m^\alpha$, with $\alpha=-2$ (top) and $\alpha=2$ (bottom). Left: total-variation distance $D_{\rm TV}(m_1)$ between the conditional mass-ratio distribution obtained from the explicit calculation and the ordered-draw prediction reconstructed from the corresponding primary-mass distribution. Curves show the three-body population before exchanges, after $N_{\rm enc}=0,$ $1$, $3$, and $10$ binary-single encounters, and after $N_{\rm enc}=3$ encounters with additional gravitational-wave merger-time weighting. Middle: corresponding residual in the effective mass-ratio slope, $\Delta\beta_q(m_1)$, between the explicit and ordered-draw distributions. Right: example conditional distributions $p(q\mid m_1)$ at $m_1=15~M_\odot$ after $N_{\rm enc}=10$, comparing the explicit calculation with the ordered-draw approximation. The small residuals show that, over most of the mass range, the effects of three-body formation and subsequent dynamical processing are largely captured by the primary-mass distribution itself, with only modest additional correlations between the two component masses.
}
\label{fig:ordered_draw_validation}
\end{figure*}

For each  model, we compare the conditional distribution $p_{\rm mod}(m_2\mid m_1)$ with the ordered-draw prediction reconstructed from that model primary-mass distribution. As in the main text, we quantify the difference using the total-variation distance
\begin{equation}
D_{\rm TV}(m_1)
=
\frac{1}{2}
\int
\left|
p_{\rm mod}(m_2\mid m_1)
-
p_{\rm od}(m_2\mid m_1)
\right|dm_2 .
\end{equation}
We also compute an effective mass-ratio slope $\beta_q(m_1)$ by fitting
\begin{equation}
p(q\mid m_1)\propto q^{\beta_q(m_1)}
\end{equation}
over the same range of $q$ for both the explicit model and the approximation.

In Fig.~\ref{fig:ordered_draw_validation} we show the results of the comparison test.
For the models considered here, the ordered-draw approximation reproduces the no-exchange three-body model well. This occurs because the dominant mass dependence of the three-body formation rate is nearly separable in the two binary components. Including binary-single exchanges does not substantially change this conclusion. Exchanges mainly reshape the effective one--body mass distribution; once the ansatz is reconstructed from $p_{1,\rm ex}(m_1)$, the residual differences in $p(m_2\mid m_1)$ remain small. Similarly, merger-time weighting mainly changes the primary-mass distribution and produces only small residual deviations from the ordered-draw relation.

We note that the residuals shown in Fig.~\ref{fig:ordered_draw_validation} are largest near the edges of the allowed mass range. This is
expected. Close to the lower boundary, the allowed interval in mass
ratio is narrow, since $q$ is restricted to $m_{\rm lo}/m_1\leq q\leq 1$.
The conditional distribution is therefore normalized over a small range, so
even modest differences in the joint distribution can appear amplified. A
similar effect occurs near the upper mass cutoff, where the ordered-draw
prediction is sensitive to the rapid truncation of the BH mass
spectrum. Finally, near $q\simeq 1$, the ordering boundary makes the
comparison especially sensitive to any weak preference for nearly equal-mass
binaries introduced by the exchange or merger-time weighting. The larger
edge residuals therefore mostly reflect the sharp boundaries of the
underlying distribution, rather than indicating a  failure of the ordered-draw
approximation. Over the interior of the allowed mass-ratio range, the
explicit dynamical models and the ordered-draw prediction remain in close
agreement.

A simplification of the preceding analytical prescriptions is that they neglect
correlations between the component masses and the semimajor axis. To assess the
importance of this approximation, we performed a controlled Monte Carlo
calculation in which the binary binding energy is tracked through successive
encounters\footnote{We thank Barry Ginat for suggesting this calculation}. We initialize binaries from the same three-body mass kernel used
above, assign an initial semimajor axis at fixed binding energy, and then evolve
the population through a prescribed number of encounters. In each encounter we
draw an interloper from the mass-dependent encounter kernel, choose the outgoing
binary using the hard-binary exchange weights, and update the semimajor axis
according to the change in the binary mass product. Specifically, after an
encounter the cutoff semimajor axis is
$
a_{\rm co}=a_{\rm in}\frac{m_{\rm out,1}m_{\rm out,2}}
{m_{\rm in,1}m_{\rm in,2}},
$
and the post-encounter value is drawn from
\(p(a)\propto a^2\) below this cutoff. This propagates the correlation between
mass and binding energy through the encounter sequence.

After a number of encounters, we apply an approximate
thermal-eccentricity-averaged probability that the binary merges in isolation
within \(13\,{\rm Gyr}\) using equation~\ref{gwtime}.
 We then compare the resulting merger-weighted distribution in
the \((m_1,m_2)\) plane with the ordered-draw reconstruction obtained from the
same primary-mass distribution.

\begin{figure*}
\centering
\includegraphics[width=0.95\textwidth]
{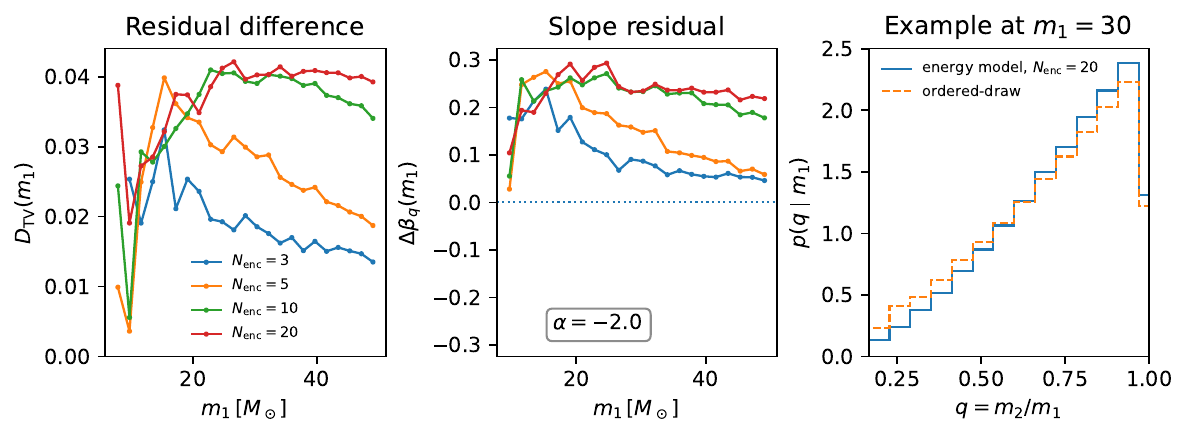}

\includegraphics[width=0.95\textwidth]
{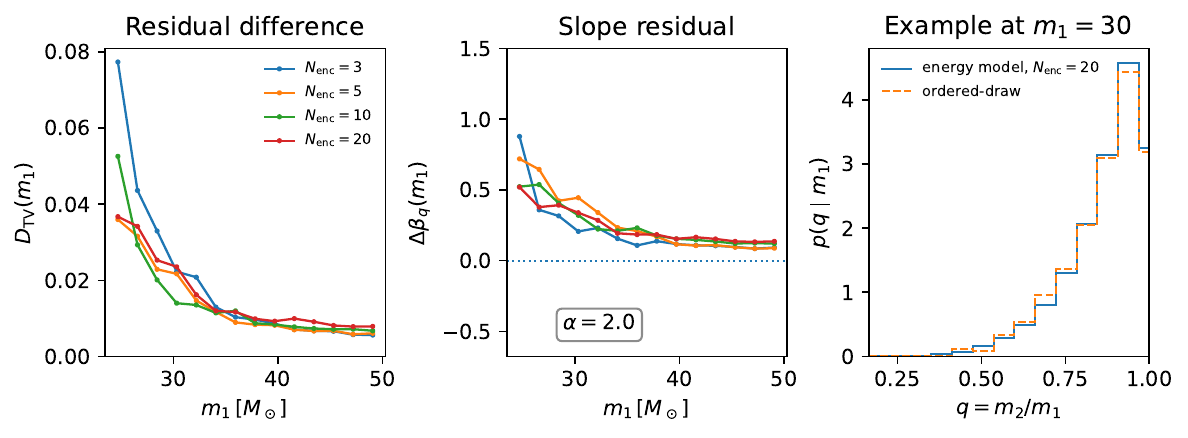}
\caption{
Same as Fig.~\ref{fig:ordered_draw_validation} but here the binary binding energy is
tracked through successive encounters. Binaries are initialized from the
three-body mass kernel and assigned semimajor axes at fixed binding energy.
After each encounter, the outgoing binary is selected using the hard-binary
exchange weights and its semimajor axis is updated according to the change in
the binary mass product. The population is then weighted by the probability of
merging in isolation within \(13\,{\rm Gyr}\), assuming a thermal eccentricity
distribution.
}
\label{fig:ordered_draw_validation_2}
\end{figure*}

Fig.~\ref{fig:ordered_draw_validation_2} shows that including this additional mass--semimajor-axis correlation
does not  degrade the ordered-draw approximation. The conditional
total-variation distance remains small, \(D_{\rm TV}\lesssim 0.05\). The
example conditional distributions \(p(q|m_1)\) are also well reproduced by the
ordered-draw reconstruction. Thus, within this controlled experiment, the
effects of binding-energy evolution and GW merger-time selection are largely
absorbed into the effective one--body mass spectrum, leaving only weak residual
pair correlations.

\subsubsection{Comparison to cluster simulations
}

In the main text we present a comparison of the ordered-draw approximation to cBHBd cluster models \cite{AntoniniGieles2020a,2023MNRAS.522..466A,2026arXiv260528088F}. Here we give more details of the comparative analysis. 

\begin{figure*}
\centering
\includegraphics[width=0.45\textwidth]
{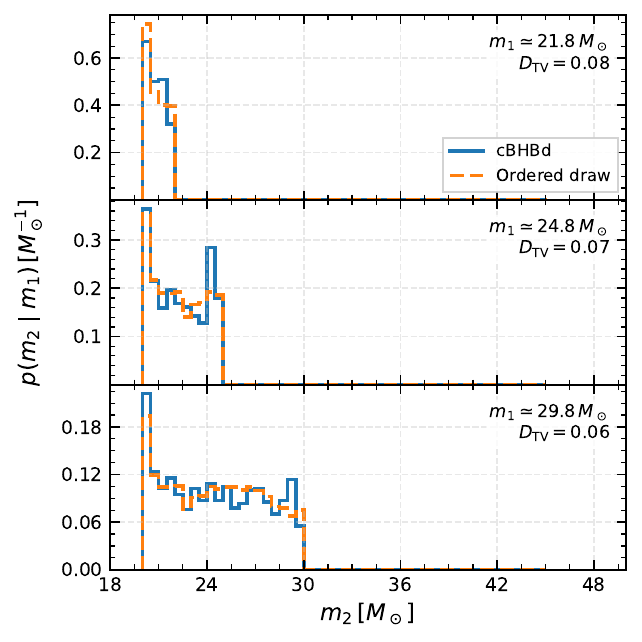}
\includegraphics[width=0.45\textwidth]
{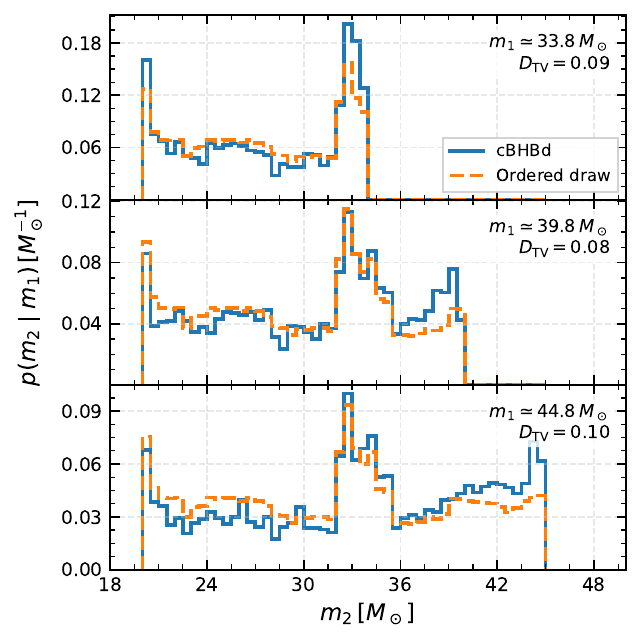}
\caption{
Each panel shows $p(m_2\mid m_1)$ at fixed primary mass, for the binary BH mergers obtained from 10,000 random realization of a cluster model  with
$N=10^5$, $\rho_{h}=10^4M_\odot~pc^{-3}$, and $Z=0.01$. The solid curves show the
cBHBd prediction, while the dashed curves show the
ordered-draw prediction reconstructed from the same primary-mass distribution.
}
\label{fig:1clpm2}
\end{figure*}

In Fig.~\ref{fig:1clpm2} we show the conditional $p(m_2|m_1)$ distribution at various primary mass values for the  BH mergers obtained from 10,000 cBHBd random realization of a cluster model  with
$N=10^5$, $\rho_{h}=10^4M_\odot~pc^{-3}$, and $Z=0.01$. Clearly
the ordered-draw model
provides a good description of  the main companion-mass features, including the secondary-mass
ridge, with the largest residuals occurring in slices where the distribution
is narrow or affected by finite-sample structure in the cluster model.

\begin{figure*}
\centering
\includegraphics[width=0.46\textwidth]
{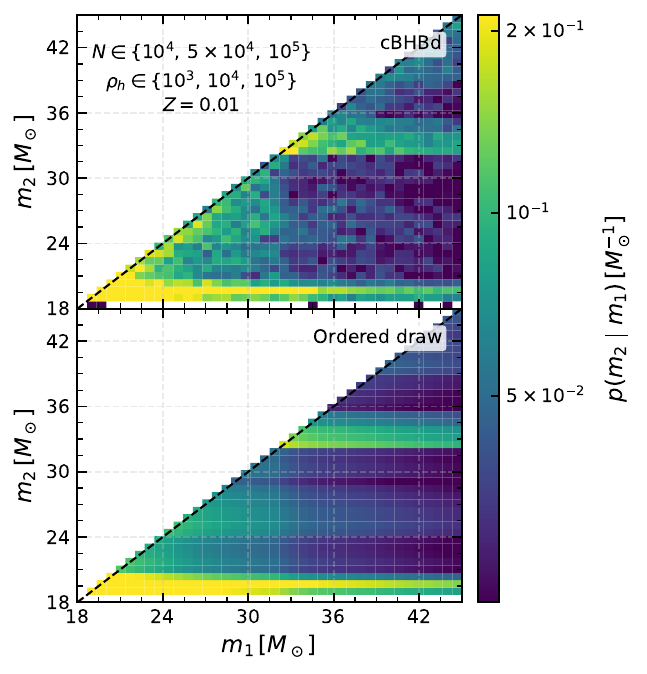}
\includegraphics[width=0.48\textwidth]
{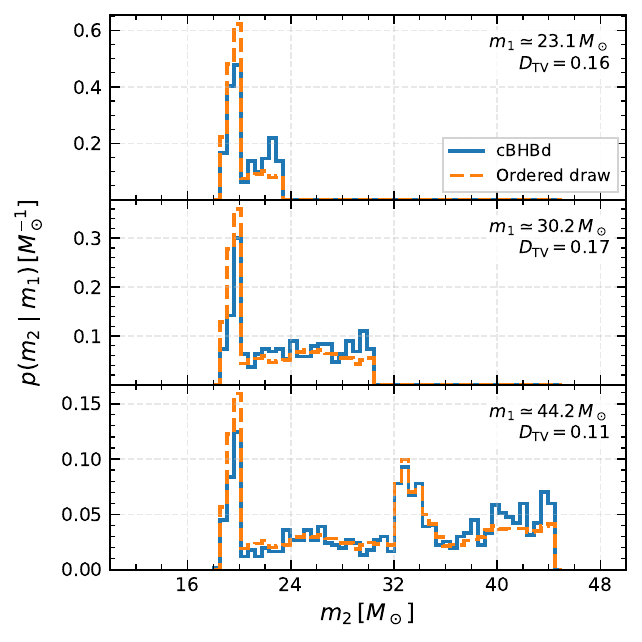}
\caption{
Left: conditional distribution of the secondary binary black-hole component masses for a representative population with three different values of cluster mass and density at a fixed metallicity. For each point in the grid, we evolved 5000 random realizations.
The upper panel shows the  distribution obtained from the cBHBd
models, while the bottom panel shows the ordered-draw prediction reconstructed
from the same primary-mass distribution. The dashed line marks the equal-mass
boundary, $m_2=m_1$; only the region $m_2\leq m_1$ is populated by
construction. The ordered-draw model reproduces the main structure of the
cluster-model distribution, including the prominent companion-mass ridges near
$m_2\simeq20\,M_\odot$ and $m_2\simeq 35\,M_\odot$, although the cluster-model histogram retains
additional finite-sample structure.
Right: Each panel shows $p(m_2\mid m_1)$ at fixed primary mass for the
cBHBd prediction and the
ordered-draw prediction reconstructed from the same primary-mass distribution.
 The values of
 $D_{\rm TV}$ quantify the difference between the two
conditional distributions in each primary-mass slice. The ordered-draw model
reproduces the main companion-mass features, including the prominent peaks seen 
in the secondary-mass
distribution.
}
\label{fig:1cljoint2}
\end{figure*}

\begin{figure*}
\centering
\includegraphics[width=0.46\textwidth]
{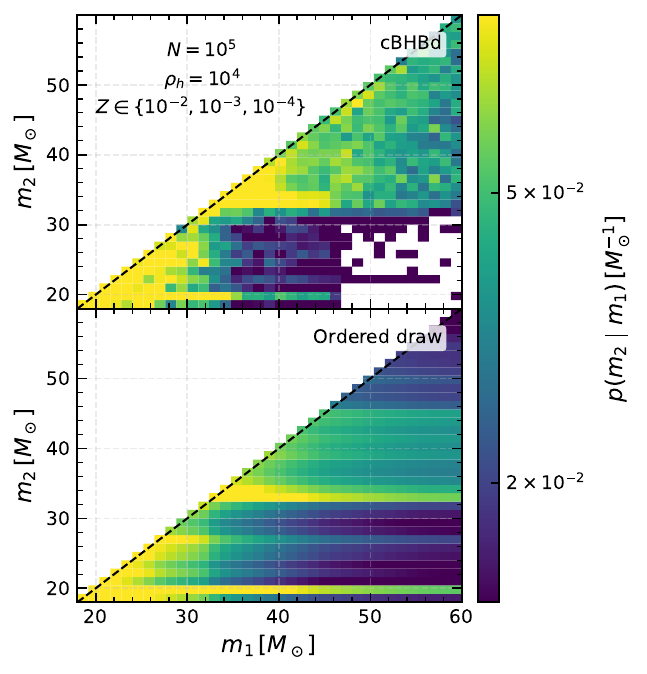}
\includegraphics[width=0.48\textwidth]
{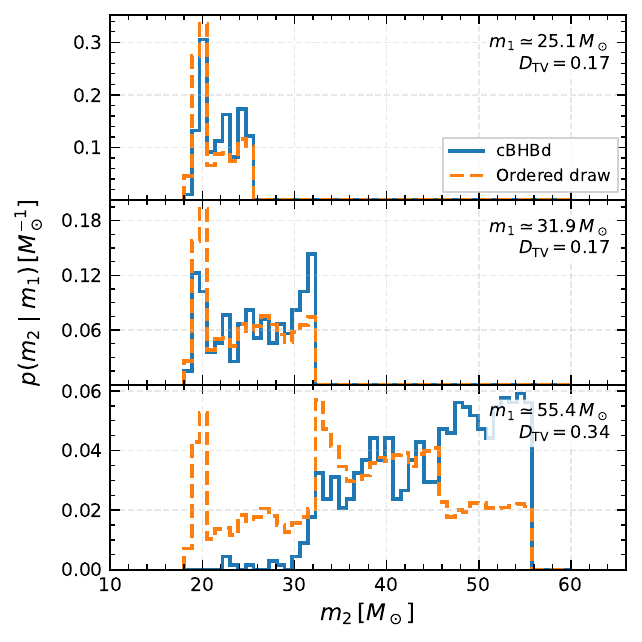}
\caption{
Same as
Fig.~\ref{fig:1cljoint2}
but for a set of models where we used three different values of cluster metallicity at a fixed mass and density.
}
\label{fig:1cljoint3}
\end{figure*}

\begin{figure*}
\centering
\includegraphics[width=0.46\textwidth]
{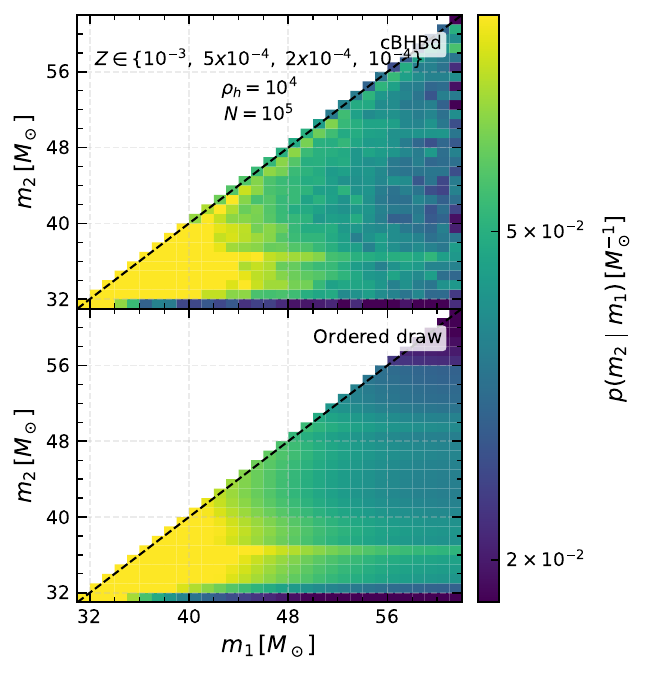}
\includegraphics[width=0.48\textwidth]
{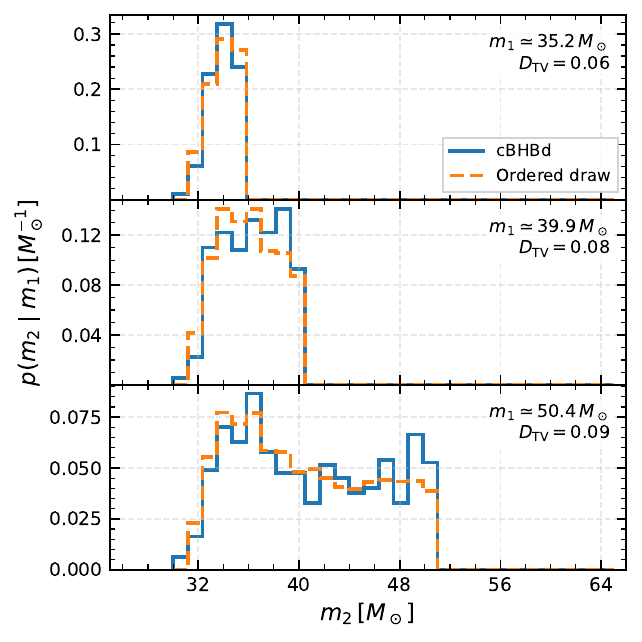}
\caption{
Same as
Fig.~\ref{fig:1cljoint2}
but for a set of models where we used four different values of cluster metallicity $Z\le0.001$ at a fixed mass and density.
}
\label{fig:1cljoint4}
\end{figure*}

Figure~\ref{fig:1cljoint2} shows the same comparison after combining models with three values of the initial cluster mass and three values of the initial cluster density, at fixed metallicity. We performed 5,000 model realizations for each combination of parameters. The cBHBd population retains a well-defined triangular support in the ordered $(m_1,m_2)$ plane and shows prominent structure at approximately fixed secondary mass. The ordered-draw prediction, reconstructed only from the primary-mass distribution of the same population, reproduces the broad support and the main secondary-mass feature. The conditional slices show residual differences,  but the overall agreement indicates that much of the pairing morphology is still controlled by the effective black-hole mass spectrum.

Figures~\ref{fig:1cljoint3} and~\ref{fig:1cljoint4} repeat the comparison for
models with different metallicities at fixed initial cluster mass and density. We find that metallicity has the larger effect.
When clusters with different metallicities are combined, sizeable residuals
appear at primary masses \(m_1\gtrsim45\,M_\odot\). These residuals are caused
by metallicity-dependent black-hole mass cutoffs. In particular, the
\(Z=0.001\) models produce no black holes above \(\simeq45\,M_\odot\), whereas
the lower-metallicity models extend to larger primary masses and mostly produce BHs with masses above \(\sim30\,M_\odot\). The combined cluster
population therefore contains an empty region at
\(m_1\gtrsim45\,M_\odot\) and \(m_2\gtrsim30\,M_\odot\). This sharp
sub-population boundary cannot be described by an aggregate primary-mass
distribution alone, so the ordered-draw reconstruction fills part of this
region.
Away from this metallicity-induced cutoff, however, the approximation still captures the shape of the distribution remarkbly well. Below \(m_1\simeq45\,M_\odot\), the ordered-draw prediction recovers the
cluster distribution with \(D_{\rm TV}\sim0.1\). Similarly, if the empty
high-\(m_2\) region is excluded from the comparison, restricting to
\(m_2\lesssim30\,M_\odot\), the agreement at high primary masses also returns
to relatively low levels \(D_{\rm TV}\lesssim0.2\). Thus, although metallicity mixing can introduce sharp cutoffs
and sub-population structure that are not fully captured by a single
ordered-draw reconstruction,  the approximation can still recover the main
features of the mass-pairing distribution.
In the second low metallicity population with   $Z\le 0.001$ the agreement is similar to the other  models with fixed metallicity across the entire mass range.
The ordered-draw model captures the broad support and the principal horizontal
features in the conditional distribution.

Finally, we  compare the ordered--draw prediction to the results of the more accurate cluster Monte Carlo models of Ref.~\cite{2019PhRvD.100d3027R}.  We consider models with metallicity $Z=0.02$, $Z=0.002$ and  with a range of initial masses, and 
virial radii $R_v$. Fig.~\ref{fig:1cljoint4} illustrates the comparison.
The predicted horizontal structures at fixed $m_2$ are clearly present in the CMC simulations, e.g., at $8M_\odot$ in the high metallcity case, and at $18M_\odot$ in the lower metallicity case which  supports the  physical picture developed in this work --at high component masses the mass ratio distribution is not peaked near one. Moreover, the right panels of Fig.~\ref{fig:1cljoint4} show that the conditional distribution at different values of primary mass is also recovered, showing that the ordered--draw model reproduces approximately  the amplitude and location of peaks in $p(m_2|m_1)$ obtained from the CMC  simulations.

\begin{figure*}
\centering
\includegraphics[width=0.46\textwidth]
{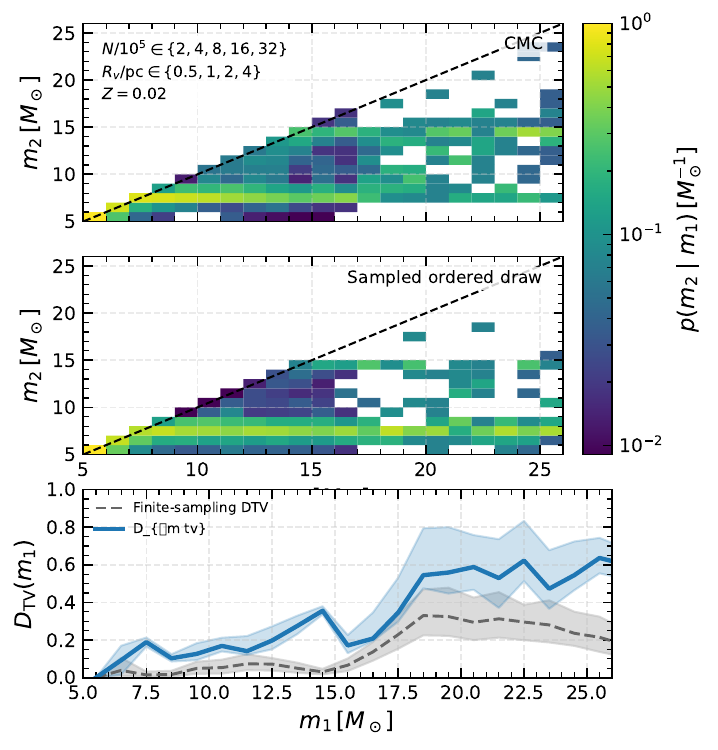}
\includegraphics[width=0.45\textwidth]
{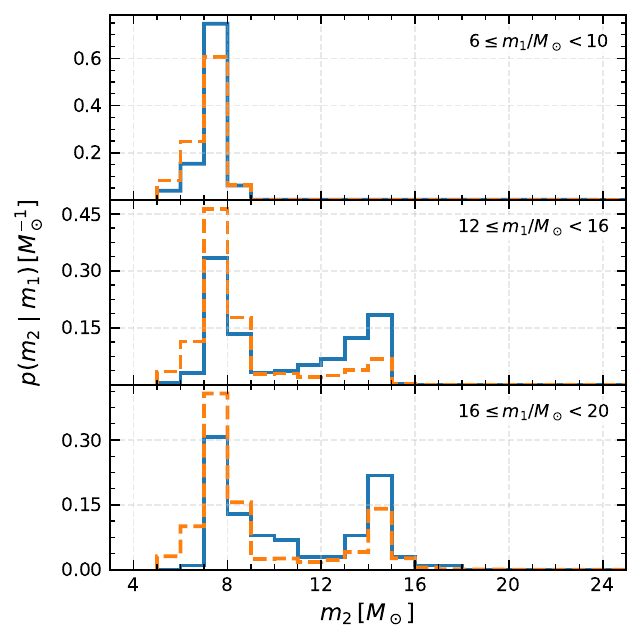}
\includegraphics[width=0.46\textwidth]
{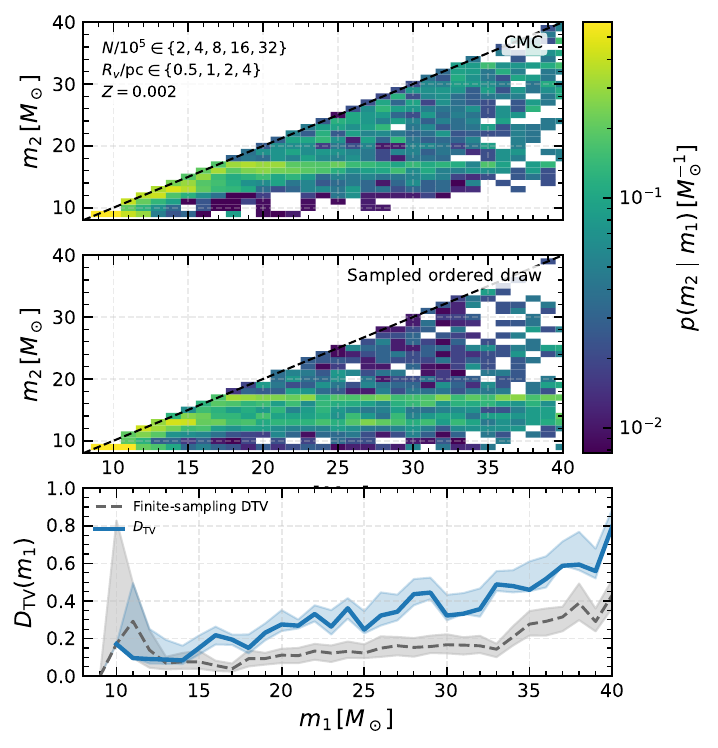}
\includegraphics[width=0.45\textwidth]
{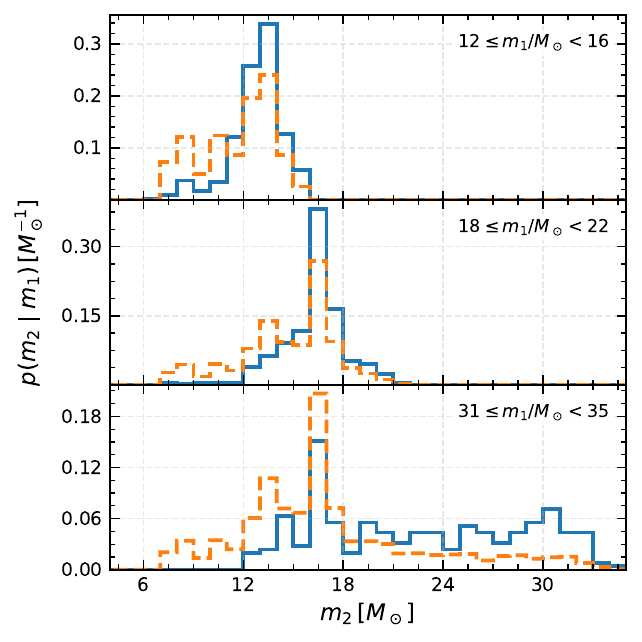}
\caption{
Comparison between the marginal secondary mass distribution $p(m_2|m_1)$ obtained from CMC models and the ordered-draw prediction. In this case, to illustrate
regions with sparse data, the ordered draw prediction is 
obtained by sampling the same number of values as in the CMC catalog of mergers. 
In the right panel we show the distribution obtained for various ranges of $m_1$, illustrating that the approximation recovers the overall structure of the simulated distribution, including the broad support at higher primary masses and  peaks. 
The grey band in the lower panel shows the 5--95 percentile range of
\(D_{\rm TV}(m_1)\) expected from finite sampling under the ordered-draw null
hypothesis. Mock catalogues are drawn from the smooth ordered-draw
reconstruction with the same number of mergers as the CMC catalogue, preserving
the number of systems in each \(m_1\) bin. The dashed grey curve shows the
median finite-sampling \(D_{\rm TV}\). 
The 
increasing discrepancies at higher masses can be  in part attributed to finite sample fluctuations due to the relatively low number of high mass mergers  in the models.
}
\label{fig:1cljoint4}
\end{figure*}

Overall, these comparisons suggest that the ordered-draw approximation captures
the leading pairing morphology of dynamically assembled binaries when the
cluster population is heterogeneous in mass and density but shares a broadly
similar black-hole mass spectrum. In this case, much of the structure in
\(p(m_2|m_1)\) can be understood as the ordered pairing of black holes drawn
from a single effective mass function, and characteristic mass scales in
that function remain visible as horizontal features in the secondary-mass
distribution. The approximation becomes less accurate when the combined
population mixes clusters with substantially different metallicities, because
the different sub-populations have different black-hole mass ranges and
high-mass cutoffs. These differences introduce sharp edges and empty regions in
the joint distribution that cannot be reconstructed from a single aggregate
primary-mass distribution alone. Nevertheless, even in this case, ordered
drawing remains a useful baseline: if the underlying black-hole mass function is
broad enough, some of the secondary-mass structure is still expected to survive,
while deviations from the ordered-draw prediction can be used to identify where metallicity
mixing and component mass correlations become important.

\section{Hierarchical mergers}

In dense star clusters, black holes formed in previous mergers can
be retained and merge again  populating an initial mass gap in the BH mass spectrum \cite{2016ApJ...824L..12O,2016ApJ...831..187A,2017PhRvD..95l4046G,Fishbach_2018}. If retention is sufficiently rare, a
cluster may contain at most one second-generation black hole within the gap at a
time \cite[although see][]{2026arXiv260407456G}. in this case,  when a remnant
participates in a binary, its companion must come from
the first-generation population. This introduces a pairing
constraint beyond mass ordering since the two components are no longer
independent draws from a single common reservoir. Consequently, the
ordered-draw approximation formally breaks down. Here we show, however, that 
hierarchical mergers do not necessarily invalidate the ordered-draw
picture.

In our approximation, the effective one-body distribution
$p_I(m)$ describes the BHs participating in mergers, including
both first-generation black holes and remnants of previous mergers.
The tendency of second-generation remnants to appear as primaries
already follows from their larger masses and the ordering condition.
A departure arises if their companion preferences cannot be explained
by mass ordering alone. For example, a departure  arises if second-generation primaries
paired only with first-generation companions, excluding other
second-generation remnants even when these satisfy $m_2<m_1$.
Such a restriction would depend on generation rather than mass
ordering alone.

To illustrate why this restriction need not substantially alter the
ordered-draw prediction, we write the effective reservoir as
\begin{equation}
p_I(m)=(1-\epsilon)p_{\rm 1G}(m)+\epsilon p_{\rm 2G}(m),
\end{equation}
where both components are normalized and $\epsilon$ is the fraction
of the effective reservoir contributed by second-generation remnants.
Defining their cumulative distributions as $F_{\rm 1G}$ and
$F_{\rm 2G}$, for a primary above the first-generation mass range ordered draw predicts
\begin{equation}
p_{\rm OD}(m_2\mid m_1)=
\frac{(1-\epsilon)p_{\rm 1G}(m_2)+\epsilon p_{\rm 2G}(m_2)}
{F_{\rm 1G}(m_1)(1-\epsilon)+\epsilon F_{\rm 2G}(m_1)}\simeq \frac{(1-\epsilon)p_{\rm 1G}(m_2)+\epsilon p_{\rm 2G}(m_2)}
{(1-\epsilon)+\epsilon F_{\rm 2G}(m_1)},
\end{equation}
since $F_{\rm 1G}(m_1)\simeq 1$ above the first-generation population.
Expanding in $\epsilon$, we obtain
\begin{equation}
\begin{aligned}
p_{\rm OD}(m_2\mid m_1)
\simeq {}& p_{\rm 1G}(m_2)&+\epsilon\left[
p_{\rm 2G}(m_2)-F_{\rm 2G}(m_1)p_{\rm 1G}(m_2)
\right]
+\mathcal{O}(\epsilon^2).
\end{aligned}
\end{equation}

Thus, when $\epsilon\ll1$, the ordered-draw approximation already
predicts a companion distribution dominated by the first-generation
population, reducing to $p_{\rm 1G}(m_2)$ as $\epsilon\to0$.
Under the assumption that companions are otherwise drawn from
$p_{\rm 1G}$, restricting hierarchical primaries to first-generation
companions therefore changes the prediction only at order $\epsilon$.
This does not mean that joint distribution can be factorized.
But it shows that hierarchical mergers  preserve the leading-order
companion distribution when the effective reservoir is dominated by
first-generation black holes. 

An important consequence of the above is that a mass gap repopulated by hierarchical
mergers need not strongly affect the secondary-mass distribution of binaries
with primaries in the gap. The steeper high-mass decline of \(p(m_2)\)
relative to \(p(m_1)\) should therefore not, by itself, be interpreted as
evidence for second-generation black holes. This behaviour follows generically
from mass ordering: \(m_1\) is the maximum of the two component masses, whereas
large values of \(m_2\) require both components to be massive.

\clearpage

\twocolumngrid

%

\end{document}